%% file: ms.tex
\documentclass{article}
\usepackage{amsmath}
\usepackage{amssymb}
\usepackage{graphicx}
\usepackage{listings}
\usepackage{enumitem}
\usepackage{algorithm, algpseudocode}
\usepackage{float}
\usepackage{bm}
\usepackage{subfig}
\usepackage{amsmath,amssymb,mathrsfs}
\usepackage{bbm}
\usepackage{url}
\usepackage{wrapfig}
\usepackage{multirow}
\usepackage[normalem]{ulem}
\usepackage{booktabs}

\usepackage[hidelinks]{hyperref}

\usepackage[round]{natbib}

\usepackage{etoolbox}
\usepackage{url}
\usepackage[most]{tcolorbox}
\usepackage{xcolor}
\input{macro_wnchen}

\usepackage{enumitem }

\newcommand*{\horzbar}{\rule[.5ex]{2.5ex}{0.5pt}}

\newtheorem{theorem}{Theorem}[section]

\newtheorem{lemma}[theorem]{Lemma}
\newtheorem{corollary}[theorem]{Corollary}

\newtheorem{definition}[theorem]{Definition}

\newtheorem{remark}[theorem]{Remark}

\title{Improved Communication-Privacy Trade-offs in $L_2$ Mean Estimation under Streaming Differential Privacy}
\author{Wei-Ning Chen$^{1}$\thanks{Part of the work was done during Wei-Ning's internship at Google. The paper has been accepted to ICML 2024.} \and Berivan Isik$^{1}$ \and Peter Kairouz$^{2}$ \and Albert No$^{3}$ \and Sewoong Oh$^{2, 4}$ \and Zheng Xu$^{2}$} 
\date{
Stanford University$^{1}$, Google Research$^{2}$, Yonsei University$^{3}$, University of Washington$^{4}$ 
} 

\begin{document}
\maketitle

\input{sec0_abstract}
\input{sec1_introduction}
\input{sec2_prior_works}

\input{sec3_preliminaries}
\input{sec4_mean_estimation}
\input{sec5_matrix_factorization}
\input{sec6_experiments}
\input{sec7_conclusion}

\newpage
\bibliographystyle{plainnat}
\bibliography{references}

\newpage
\input{sec99_appendix}
\end{document}

%% file: macro_wnchen.tex
\newcommand{\qedwhite}{\hfill \ensuremath{\Box}}

\newcommand{\lp}{\left(}
\newcommand{\rp}{\right)}
\newcommand{\lb}{\left[}
\newcommand{\rb}{\right]}
\newcommand{\lbp}{\left\{}
\newcommand{\rbp}{\right\}}
\newcommand{\lba}{\left\lvert}
\newcommand{\rba}{\right\rvert}
\newcommand{\lV}{\left\lVert}
\newcommand{\rV}{\right\rVert}

\newcommand{\mcal}{\mathcal}

\newcommand{\mb}{\mathbf}
\newcommand{\bbm}{\mathbbm}
\newcommand{\mbb}{\mathbb}
\newcommand{\msf}{\mathsf}

\newcommand{\ra}{\rightarrow}

\newcommand{\eqDef}{\triangleq}
\newcommand{\diid}{\overset{\text{i.i.d.}}{\sim}}

\newcommand{\E}{\mathbb{E}}

%% file: sec0_abstract.tex
\begin{abstract}
    We study $L_2$ mean estimation under central differential privacy and communication constraints, and address two key challenges: firstly, existing mean estimation schemes that simultaneously handle both constraints are usually optimized for $L_\infty$ geometry and rely on random rotation or Kashin's representation to adapt to $L_2$ geometry, resulting in suboptimal leading constants in mean square errors (MSEs); secondly, schemes achieving order-optimal communication-privacy trade-offs do not extend seamlessly to streaming differential privacy (DP) settings (e.g., tree aggregation or matrix factorization), rendering them incompatible with DP-FTRL type optimizers.
    In this work, we tackle these issues by introducing a novel privacy accounting method for the sparsified Gaussian mechanism that incorporates the randomness inherent in sparsification into the DP noise. Unlike previous approaches, our accounting algorithm directly operates in $L_2$ geometry, yielding MSEs that fast converge to those of the uncompressed Gaussian mechanism. Additionally, we extend the sparsification scheme to the matrix factorization framework under streaming DP and provide a precise accountant tailored for DP-FTRL type optimizers. Empirically, our method demonstrates at least a 100x improvement of compression for DP-SGD across various FL tasks. 

    
\end{abstract}
\looseness=-1

%% file: sec1_introduction.tex
\section{Introduction}
In federated learning (FL) \citep{mcmahan2016communication, konevcny2016federated, kairouz2021advances}, a server executes a specific learning task on data that is kept on clients' devices, avoiding the explicit collection of local raw datasets. This process typically involves the server iteratively gathering essential local model updates (such as noisy gradients) from the client side and subsequently updating the global model. While FL embodies the principle of data minimization by only requesting the minimal information necessary for model training, these local model updates may still contain sensitive information. As a result, additional privacy protection is necessary to prevent the trained model from possibly revealing individual information. Moreover, with the increase of model size, the exchange of local model updates becomes both memory and computation-intensive, leading to substantial latency and impeding the efficiency of training cycles. Consequently, it is desired to devise robust privacy protection mechanisms that simultaneously optimize communication efficiency.

\looseness=-1
In this paper, we study the $L_2$ mean estimation\footnote{Here, $L_2$ refers to the $L_2$ geometry of the local model updates, i.e., $\lV \bm{g}_i \rV_2 \leq \Delta_2$ for all client $i$. This condition is typically maintained through the $L_2$ clipping step of the differential privacy mechanism.}, a core sub-routine in the majority of FL schemes, subject to joint communication and differential privacy (DP)~\citep{dwork2006calibrating} constraints. We consider two major types of DP optimization settings: (1) the classic DP-SGD type approach \citep{abadi2016deep} where independent DP noise is injected in each round of training, and (2) the DP-FTRL type approach \citep{guha2013nearly, kairouz2021practical, denisov2022improved} where the DP noise is correlated across training rounds, the structure of which is intricately designed based on certain matrix factorization.

\looseness=-1
There has been a long thread of literature on distributed mean estimation (DME) under either or both privacy and communication constraints \citep{an2016distributed, konevcny2016federated, agarwal2018cpsgd, chen2020breaking, chen2023privacy, shah2022optimal, feldman2021lossless, isik2023exact, asi2023fast}. Recent work by \citet{chen2023privacy} points out that, to achieve order-optimal mean square errors (MSEs) under joint constraints, it becomes imperative to integrate the inherent randomness utilized in compression (e.g., in sampling, sketching, or projection) into privacy analysis. Essentially, the implicit ``compression noise'' should be leveraged to amplify the DP guarantees, resulting in a significant reduction of DP noise. However, despite the coordinate subsampled Gaussian mechanism (CSGM) introduced by \citet{chen2023privacy} achieving order-optimal MSEs, it is crafted within the $L_\infty$ geometry (i.e., assuming $\lV \bm{g}_i\rV_\infty \leq C_\infty$ for any local vector $\bm{g}_i$) and relies on random rotation or Kashin's representation to extend to $L_2$ mean estimation tasks. It is noteworthy that the bounded $L_2$ norm assumption is \emph{strictly more robust} than the bounded $L_\infty$ norm assumption (see Section~\ref{subsec:l_inf_comparison}), inevitably leading to larger MSEs with CSGM compared to the (uncompressed) Gaussian mechanism under equivalent DP guarantees.

\looseness=-1
A further challenge arises in CSGM (or, more broadly, general randomized compression schemes based on random projection, sampling, or sketching) when applied to streaming DP models \citep{guha2013nearly, denisov2022improved,  jain2023price}, particularly in the context of DP-FTRL type optimization mechanisms based on tree aggregation \citep{honaker2015efficient, kairouz2021practical} or matrix factorization \citep{denisov2022improved}. In the streaming DP model, the DP noise injected in each round loses its independence. Instead, noise variables $\mb{Z} \in \mbb{R}^T$ are correlated across $T$ training rounds through a linear transform $\mb{B}\cdot\mb{Z}$, where $\mb{B} \in \mbb{R}^{T\times T}$ is obtained from certain matrix factorization of the objective function that aims to minimize the overall distortions, such as MSEs. When the noise variables are correlated across rounds, they are no longer ``aligned'' with the randomness introduced in the local compression phase, as compression occurs locally and is thus independent across rounds. This complicates the analysis of privacy amplification, as privacy budgets cannot be accounted for round-wise, introducing what we term ``temporal coupling.'' Moreover, the adaptive nature of DP-FTRL, where local gradients depend on the outputs of previous rounds, leads to the coupling of compression seeds that are typically introduced independently across dimensions. When analyzing the outputs over $T$ rounds, this coupling, referred to as ``spatial coupling,'' presents a significant challenge. Traditional privacy amplification tools \citep{balle2018privacy, zhu2019poission, wang2019subsampled} fail in the face of such spatial and temporal coupling, necessitating a novel analysis approach.

\looseness=-1
\paragraph{Our contribution.} In this work, we tackle both aforementioned challenges. Firstly, we introduce a novel privacy accounting method for the sparsified Gaussian mechanism. This method incorporates the inherent randomness from the sparsification phase into the DP noise. Unlike previous approaches in \citet{chen2023privacy}, our accounting algorithm directly operates in $L_2$ geometry, resulting MSEs that converge fast to those of the uncompressed Gaussian mechanism. The key technique is to leverage the convexity of the R\'enyi DP profile of $1$-dimensional subsampled Gaussian mechanism and extend it to multi-dimensional scenarios.

\looseness=-1
Secondly, we extend the application of the sparsified Gaussian mechanism to streaming DP settings, particularly within the matrix factorization DP-FTRL framework. We establish a R\'enyi privacy accounting theorem. While this theorem bears similarities to its non-streaming counterpart, the analysis necessitates a fundamentally different approach due to the spatial and temporal coupling inherent in the adaptive releases. A crucial step in our analysis involves decomposing the transcript (i.e., the collection of all releases across $T$ training rounds), effectively transforming the adaptive releasing model into a non-adaptive one.

\looseness=-1
Although our analysis primarily revolves around the sparsified Gaussian mechanism (or coordinate subsampled Gaussian mechanism), it inherently encompasses a broader family of random projections, including subsampled randomized Hadamard transform \citep{ailon2006approximate, sarlos2006improved}, and randomized Gaussian design \citep[Section 6]{wainwright2019high}. These dimensionality reduction techniques can be viewed as a linear transform followed by a subsampling step. Additionally, by slightly lifting the dimension, these random designs exhibit deep connections to Kashin's representation, providing a uniform $L_\infty$ bound, albeit with a larger leading constant \citep{lyubarskii2010uncertainty}.

\looseness=-1
Finally, we present comprehensive empirical results on the proposed $L_2$ sparsified Gaussian mechanism and sparsified Gaussian matrix factorization. Our results demonstrate a $100 \times$ improvement in compression rates in various FL tasks (including FMNIST and Stackoverflow datasets). Moreover, our algorithm reduces the dimensionality of local model updates and hence can potentially be combined with other quantization or (scalar) lossless compression techniques~\citep{alistarh17qsgd, isik2022information, mitchell2022optimizing}.

%% file: sec2_prior_works.tex
\section{Related Work}
\textbf{FL and DME.} Federated learning \citep{konevcny2016federated, mcmahan2016communication, kairouz2019advances} emerges as a decentralized machine learning framework that provides data confidentiality by retaining clients' raw data on edge devices. In FL, communication between clients and the central server can quickly become a bottleneck \citep{mcmahan2016communication}, so previous works have focused on compressing local model updates via gradient quantization \citep{mcmahan2016communication, alistarh17qsgd, g2019vqsgd, an2016distributed, wen2017terngrad, wangni2018gradient, braverman2016communication}, sparsification \citep{barnes2020rtopk, hu2021federated, farokhi2021gradient, isik2023sparse, lin2018deep}, or random projection~\citep{rothchild2020fetchsgd, vargaftik2021drive}. To further enhance user privacy, FL is often combined with differential privacy \citep{dwork2006calibrating, abadi2016deep, agarwal2018cpsgd, hu2021federated}.

\looseness=-1
Note that in this work, we consider FL (or more specifically, DME) under a \emph{central}-DP setting where the server is trusted, which is different from the local DP model \citep{kasiviswanathan2011can}\footnote{Another alternative to private DME is via local DP and shuffling. We provide a detailed discussion on this direction in Appendix~\ref{sec:for_apple_folks}} and the distributed DP model with secure aggregation \citep{bonawitz2016practical, bell2020secure, kairouz2021distributed, agarwal2021skellam, chen2022poisson, chen2022fundamental}. When the secure aggregation is employed, local model updates cannot be compressed independently \citep{rothchild2020fetchsgd, chen2023privacy}, and hence, the corresponding compression rates must be strictly higher than those without secure aggregation.

\looseness=-1
\textbf{Streaming DP and DP-FTRL.}
\looseness=-1
In addition to the classic DP optimizers such as DP-SGD \citep{abadi2016deep} or DP-FedAvg \citep{mcmahan2016communication}, we also study the online optimization settings such as DP-FTRL \citep{kairouz2021practical} where the noise is correlated across rounds. This is motivated by the facts that (1) subsampling is often impractical in federated learning settings \citep{kairouz2021practical, kairouz2019advances}, and (2) the correlated noise probably yields better utility compared to the independent noise \citep{choquette2023correlated}. The key component behind the DP-FTRL relies on the private releases under continual observation, an old problem dating back to \citep{dwork2010differential, chan2012differentially}. Since then, several works have studied the continual release model and its applications \citep{upadhyay2021framework, choquette2022multi, choquette2023correlated, henzinger2023almost, henzinger2024unifying}. \citet{kairouz2021practical} originally used the efficient DP binary-tree estimator \citep{honaker2015efficient} for the DP-FTRL algorithm, but later, a more general approach to cumulative sums based on matrix factorization \citep{hardt2010geometry, li2015matrix, yuan2016optimal, mckenna2018optimizing, edmonds2020power} was used. We, however, note that DP online optimization concerns \emph{adaptive} inputs, that is, the future data points depend on previous outputs, and not all matrix mechanisms extend to the adaptive settings \citep{denisov2022improved}, and it introduces challenges when incorporating compression into the privacy analysis. Indeed, to prove the adaptive DP guarantees of our algorithm, we need to handle the spatial and temporal dependency carefully. Finally, while the recent work \citet{choquette2023privacy} also investigate privacy amplification through subsampling, their subsampling is conducted client-wise rather than coordinate-wise, as their scheme is not designed for compression. Consequently, \citet{choquette2023privacy} do not encounter the spatial coupling issue as we do.

%% file: sec3_preliminaries.tex
\section{Preliminaries and Setups}
In this section, we introduce the mathematical formulation of the problem and the DP models. We begin with DME in non-streaming DP, and then transition to the continual sum (or mean) problem within the streaming DP model.


\subsection{DME and (Non-streaming) DP}\label{sec:dme_non_streaming}
Consider $n$ clients, each with a local vector (e.g., local gradient or model update) $\bm{g}_i \in \mathbb{R}^d$ that satisfies $\lV \bm{g}_i \rV_2 \leq \Delta_2$ for some constant $\Delta_2>0$ (one can think of $\bm{g}_i$ as a clipped local gradient).  A server aims to learn an estimate $\hat{\mu}$ of the mean $\mu(\bm{g}^n) \triangleq \frac{1}{n}\sum_i \bm{g}_i$ from $\bm{g}^n = (\bm{g}_1, \dots, \bm{g}_n)$ after communicating with the $n$ clients. Toward this end, each client locally compresses $\bm{g}_i$ into a $b$-bit message $Y_i = \mcal{E}_i\lp \bm{g}_i \rp \in\mcal{Y}$ through a local encoder $\mcal{E}_i: \mbb{R}^d \mapsto \mcal{Y}$ (where $\lba \mcal{Y} \rba\leq 2^b$) and sends it to the central server, which upon receiving $Y^n = \lp Y_1,\dots,Y_n\rp$ computes an estimate $\hat{\mu}\lp Y^n\rp$ that satisfies the following differential privacy:
\begin{definition}[Differential Privacy \citep{dwork2006calibrating}] \label{def:dp}
A mechanism (i.e., a randomized mapping) $\mcal{M}(\bm{g}^n)$ is $(\varepsilon, \delta)$-DP if for any neighboring datasets $\bm{g}^n \eqDef (\bm{g}_1,..., \bm{g}_i, ... ,\bm{g}_n)$, $\bm{h}^n \eqDef (\bm{g}_1,..., \bm{g}_{i-1}, \bm{h}_i, \bm{g}_{i+1}, ..., \bm{g}_n)$, and measurable $\mcal{S} \in  \msf{range}\lp \mcal{M} \rp$, it holds that
\begin{align*}
    \Pr\lbp \mcal{M}(\bm{g}^n) \in \mcal{S}\rbp \leq e^\varepsilon\cdot   \Pr\lbp \mcal{M}(\bm{h}^n) \in \mcal{S} \rbp + \delta,
\end{align*}
where the probability is taken over the randomness of $\mcal{M}(\cdot)$. 
\end{definition}
Our goal is to design schemes that minimize the MSE:
\begin{align*}
    \min_{\lp \mcal{E}_1, \dots, \mcal{E}_n, \hat{\mu}\rp} \max_{\bm{g}^n} \E\lb \lV \hat{\mu}\lp \mcal{E}_1(\bm{g}_1),... , \mcal{E}_n(\bm{g}_n) \rp - \mu(\bm{g}^n) \rV^2_2 \rb,
\end{align*}
subject to $b$-bit communication and $(\varepsilon, \delta)$-DP constraints.

\looseness=-1
The above DME task is closely related to FL with batched SGD (or other similar stochastic optimization methods, such as FedAvg \citep{mcmahan2016communication}). \emph{In each round}, the server updates the global model using a noisy mean of local model updates. This estimate is typically derived through a DME primitive. As demonstrated in \citep{ghadimi2013stochastic}, if the estimate remains unbiased in each round, convergence rates depend on the $L_2$ estimation error. Note that the DME procedure is invoked independently in each round, and the privacy budget is allocated for $T$ rounds of training, differing from the online DP setting discussed below.

\looseness=-1
\subsection{Streaming Differential Privacy}\label{subsec:streaming_dp}
Next, we introduce the streaming DME problem and matrix mechanisms. To begin with, we first summarize the streaming DP setting \citep{denisov2022improved}.  A streaming mechanism $\mcal{M}$ takes inputs $\bm{g}^{(1)}, \bm{g}^{(2)},...,\bm{g}^{(t)}$ and outputs $\bm{o}^{(t)}$ at time $t$. We denote the stream with $T$ rounds in the following matrix form:
$$ \mb{G} \eqDef \begin{bmatrix}
    \horzbar \, \bm{g}^{(1)} \, \horzbar\\
    \vdots \\
    \horzbar \,  \bm{g}^{(T)} \, \horzbar
\end{bmatrix}, $$
and similarly for $\mb{H}$ and the adversary's view $\mb{O}$.

\looseness=-1
An adversary that adaptively defines two input sequences $\mb{G} = (\bm{g}^{(1)}, ..., \bm{g}^{(T)})$ and $\mb{H} = (\bm{h}^{(1)},...,\bm{h}^{(T)})$. The adversary must satisfy the promise that these sequences correspond to neighboring data sets. The privacy game proceeds in rounds. At round $t$, the adversary generates $\bm{g}^{(t)}$ and $\bm{h}^{(t)}$, \emph{as a function of} $\bm{o}^{(1)},...,\bm{o}^{(t-1)}$. The game accepts these if the input streams defined so far are valid, meaning that there exist completions $\tilde{\mb{G}} = \lp\bm{g}^{(1)}, ..., \bm{g}^{(t)}, \tilde{\bm{g}}^{(t+1)},...,\tilde{\bm{g}}^{(T)}\rp$ and $\tilde{\mb{H}} = \lp\bm{h}^{(1)}, ..., \bm{h}^{(t)}, \tilde{\bm{h}}^{(t+1)},...,\tilde{\bm{h}}^{(T)}\rp$ such that $\tilde{\mb{G}}$ and $\tilde{\mb{H}}$ are neighboring, in the following sense:
\begin{definition}[Neighboring datasets]\label{def:streaming_neighboring}
    Two data streams $\mb{G}$ and $\mb{H}$ in $\mbb{R}^{T\times d}$ will be considered to be neighboring if they differ by a single row, with the $\ell_2$-norm of the difference in this row at most $\Delta_2$.
\end{definition}

\looseness=-1
The game is parameterized by a bit $\msf{side}\in\{0, 1\}$, which is unknown to the adversary but constant throughout the game. The game hands either $\mb{G}$ or $\mb{H}$ to $\mcal{M}$, depending on $\msf{side}$. We say $\mcal{M}$ is $(\alpha, \varepsilon(\alpha))$ R\'enyi DP if the adversary's views $\mb{O}$ under $\msf{side} = 0$ and $\msf{side} = 1$ is $\varepsilon(\alpha)$-indistinguishable under R\'enyi divergence at order $\alpha$:
$ D_\alpha\lp P_{\mb{O}|\mb{G}} \middle\Vert P_{\mb{O}|\mb{H}}  \rp \leq \varepsilon(\alpha).$

\looseness=-1
\subsection{DME and Matrix Mechanisms}\label{subsec:matrix_mechanism}
Finally, we consider DME under the streaming DP model. In each round $t$, the server selects a batch of clients $B_t \in [N]$ and computes the empirical mean of their local vectors $\bm{g}^{(t)} = \sum_{i \in B_t} \bm{g}_i$. Note that $\bm{g}_i$ can depend on previous outputs $\bm{o}^{(1)}, ..., \bm{o}^{(t-1)}$. Our scheme assumes single-participation-per-epoch, that is, $B_t$ disjoint with $B_{t'}$. 

\looseness=-1
The goal of matrix mechanisms is to continually release a differentially private version of $\mb{A}\mb{G}$ while minimizing the overall MSE: $\lV \widehat{\mb{A}\mb{G}} - \mb{A}\mb{G} \rV_F^2$. Here, $\mb{A} \in \mbb{R}^{T\times T}$ must be a lower triangular matrix in order to ensure causality. In online optimization, the matrix $\mb{A}$ is determined by update rules. For instance, in simple SGD with a fixed step size $\eta > 0$, the model is updated as follows:
$$ \bm{w}^{(t)} = \bm{w}^{(t-1)} - \eta \bm{g}^{(t)} = \bm{w}^{(0)} - \eta \sum_{\tau=1}^t\bm{g}^{(\tau)},$$
resulting in the corresponding $\mb{A}$ being the prefix-sum matrix satisfying  $[\mb{A}]_{t, t'} = \bbm{1}_{\{t\leq t'\}}$. In general, one can leverage the matrix mechanism within the DP-FTRL framework \citep[Algorithm~1]{kairouz2021practical} and further incorporate momentum \citep{denisov2022improved}.

\looseness=-1
To ensure privacy, instead of directly privatizing data matrix $\mb{G}$ (which results in a DP-SGD type scheme), we leverage the factorization $\mb{A} = \mb{B}\mb{C}$ for $\mb{B}, \mb{C} \in \mbb{R}^{T\times T}$. If $(\alpha, \varepsilon(\alpha))$-DP is preserved for $\mb{C}\mb{G} + \mb{Z}$, then the same level of DP holds for $\mb{A}\mb{G} + \mb{B}\mb{Z}$ as well. Notbaly, in the online optimization setting, local vectors $\bm{g}^{(t)}$ are \emph{adaptively} generated and depend on $(\bm{o}^{(1)},...,\bm{o}^{(t-1)})$. \citet[Theorem~2.1]{denisov2022improved} shows that for Gaussian mechanism (i.e., $[\mb{Z}]_{t, j} \diid \mcal{N}(0, \sigma^2)$ for some $\sigma^2 > 0$), the non-adaptive DP guarantee (meaning that $\mb{G}$ is independent with the previous private outputs $\mb{O}$) implies the same level of adaptive DP.

To optimize the error, \citet{li2015matrix, yuan2016optimal, denisov2022improved} formulate the factorization $\mb{A} = \mb{B}\mb{C}$ as a convex optimization problem :
\begin{align}\label{eq:dp_factorization_problem}
    \min_{\mb{B}:\mb{A=BC}, \, \Delta\lp\mb{C}\rp = 1} \lV \mb{B} \rV_F^2,
\end{align} 
where $\Delta\lp\mb{C}\rp \eqDef \max_{t \in [T]} \lV \mb{C}_{[:, t]} \rV^2_2$ is the sensitivity of $\mb{C}$. In this work, while we plug in the optimal factorization in our scheme (specifically solved via the fixed point method in \citet{denisov2022improved}), our results hold for general factorization.

Our objective is to devise a local compression mechanism satisfying two criteria: (1) $\widehat{\mb{A}\mb{G}}$ satisfies adaptive streaming DP, and (2) $\widehat{\mb{A}\mb{G}}$ is a function of locally compressed vectors $\mcal{E}_i\lp \bm{g}_i \rp$ that can be described in $b$ bits.

\begin{remark}
In the streaming scenario, the cohort size $|B_t|$ solely impacts the sensitivity of the mean function each round. For simplicity in privacy analysis, we assume $|B_t| = 1$ (non-batched SGD). Nevertheless, our results extend to general batch sizes, as outlined in the main theorems.
\end{remark}

\paragraph{Notation.} In the non-streaming setting, we employ $\bm{g}_i$ (or $\bm{h}_i$) to represent the local (row) vector at client $i$. In the streaming scenario, $\bm{g}^{(t)}$ (or $\bm{h}^{(t)}$) denotes the averaged row vectors of clients at round $t$. Matrices are denoted by capital bold-faced symbols; for instance, $\mb{G} \in \mbb{R}^{T\times d}$ represents the matrix form of the stream $(\bm{g}^{(1)}, ..., \bm{g}^{(T)})$, where the $t$-th row of $\mb{G}$ is $\bm{g}^{(t)}$. When the context is clear, we may use $\mb{G}$ to refer to the stream itself. Additionally, we use $\bm{g}^{(t)}_j$ or $\mb{G}_{t,j}$ interchangeably to indicate the $(t, j)$-th entry of $\mb{G}$, with $t \in [T]$ and $j \in [d]$\footnote{In general, we use $t \in [T]$ as the time index, $j \in [d]$ as the coordinate (spatial) index, and $i \in [n]$ as the client index for subscripts and superscripts.}.

%% file: sec4_mean_estimation.tex
\section{Differentially Private $\mathbf{L_2}$ Mean Estimation}
In this section, we consider the non-streaming DME problem described in Section~\ref{sec:dme_non_streaming}. To reduce communication costs under central DP, previous work of \citet{chen2023privacy} proposes coordinate-subsampled Gaussian mechanism (CSGM), which random sparsifies each local vector in a coordinate-wise fashion, followed by server aggregation and the addition of Gaussian noise. While aligning with several gradient compression techniques, CSGM significantly enhances privacy guarantees by incorporating the randomness introduced in the sparsification phase into privacy analysis.

\looseness=-1
However, a notable drawback in \citet{chen2023privacy} emerges within the $L_\infty$ geometry assumption that requires $\lV \bm{g}_i \rV_\infty \leq \Delta_\infty$. It is crucial to note that, in general, the $L_\infty$ assumption is weaker than the $L_2$ assumption described in Section~\ref{sec:dme_non_streaming}. To extend to the $L_2$ scenario, \citet{chen2023privacy} employs random rotation (or Kashin’s representation) and $L_\infty$ clipping to pre-process local vectors. This approach, however, results in larger Mean Squared Errors (MSEs) compared to the uncompressed Gaussian mechanism under equivalent Differential Privacy (DP) guarantees.

\looseness=-1
\begin{algorithm}[h]
\caption{$L_2$-CSGM}\label{alg1:CSGM}
\begin{algorithmic}
    \State {\bfseries Input:} users' data $\bm{g}_1,...,\bm{g}_n$, sampling parameters $\gamma \eqDef b/d$, DP parameters $(\alpha, \varepsilon(\alpha))$.
    \For{user $i \in [n]$}
	\State Draw $\bm{s}_{i} \diid \msf{Bern}^{\otimes d}(\gamma)$ via shared randomness.
        \State Compress and send $\bm{g}_{i} \odot \bm{s}_i$ to the server (where $\odot$ denotes the entry-wise product).
    \EndFor
    \State Server computes the noisy mean 
    \begin{align}\label{eq:CSGM}\textstyle
        \hat{\mu}_{\msf{CSGM}}(\bm{g}^n; \bm{s}^n, Z) \eqDef \frac{1}{n\gamma}\lp \sum_{i=1}^n \bm{g}_i \odot \bm{s}_i  + Z\rp,
    \end{align}
    where $Z \sim \mcal{N}\lp 0, \sigma^2\mbb{I}_d\rp$ and $\sigma^2$ is computed according to \eqref{eq:l2_rdp_bdd} in Theorem~\ref{thm:mean_estimation_l2}.
    \State {\bfseries Return:} $\hat{\mu}_{\msf{CSGM}}$.
\end{algorithmic}
\end{algorithm}

\looseness=-1
To address the geometry issue in \citet{chen2023privacy}, we introduce a slight modification (see Algorithm~\ref{alg1:CSGM}) to the CSGM scheme and present an enhanced analysis of its R\'enyi DP profile, yielding a significantly improved guarantee. To differentiate between the two schemes, we term our proposed version as $L_2$-CSGM, while the original one in \citet{chen2023privacy} is referred to as $L_\infty$-CSGM. Our main result in this section is the following privacy upper bound for the $L_2$-CSGM mean estimation scheme.
\begin{theorem}\label{thm:mean_estimation_l2}
    Let $\bm{g}_1, ..., \bm{g}_n \in \mbb{S}^{d-1}(\Delta_2)$ (i.e., $\lV \bm{g}_i \rV_2 \leq \Delta_2$), and $\lV \bm{g}_i\rV_\infty \leq \Delta_\infty$ for all $i \in [n]$. Let $\hat{\mu}_\msf{CSGM}(\bm{g}^n)$ be defined as in \eqref{eq:CSGM} with $\bm{s}_1, ...,\bm{s}_n \diid \msf{Bern}(\gamma)^{\otimes d}$, and $Z \sim \mcal{N}\lp 0, \sigma^2\mbb{I}_d\rp$. Then $\hat{\mu}_\msf{CSGM}$ satisfies
    $\lp \alpha, \varepsilon\lp \alpha \rp \rp$-R\'enyi DP, for all integer $\alpha$ and
    \begin{align}\label{eq:l2_rdp_bdd}
        \varepsilon\lp \alpha \rp \geq &\frac{ \Delta_2^2/\Delta_\infty^2}{\alpha-1} \log\Big( (1-\gamma)^{\alpha-1}\lp \gamma(\alpha-1) + 1 \rp + \nonumber\\
        &\quad \textstyle \sum_{\ell=1}^\alpha { \alpha \choose \ell}\lp 1-\gamma  \rp^{\alpha - \ell} \gamma^{\ell} e^{(\ell-1) \ell\frac{\Delta_\infty^2}{2\sigma^2}} \Big).
    \end{align}
\end{theorem}
\looseness=-1
While $L_2$-CSGM also employs $L_\infty$ clipping, we do not account for privacy budgets directly based on the $L_\infty$ clipping norm $\Delta_\infty$ (which is the case in $L_\infty$-CSGM). Instead, we consider both $\Delta_2$ and $\Delta_\infty$, with $L_\infty$ serving to ``mitigate'' the regime on which the privacy amplification lemma operates. In $L_2$-CSGM, the $L_\infty$ clipping norm only influences higher-order terms in the final guarantees, and a slight increase in $\Delta_\infty$ does not alter the privacy guarantee asymptotically with increasing dimension $d$. In the subsequent subsection, we demonstrate that, for any $\alpha > 0$ and under the same MSE constraint, the R\'enyi DP guarantee of $L_2$-CSGM converges to that of the (uncompressed) Gaussian mechanism as $d \rightarrow \infty$.

\looseness=-1
\subsection{Compared to the Gaussian Mechanism}
We begin with the following lemma that computes the MSE of $\hat{\mu}_\msf{CSGM}$.
\begin{corollary}\label{cor:CSGM_mse}
    Under the hypotheses of Theorem~\ref{thm:mean_estimation_l2}, let $\hat{\mu}_\msf{CSGM}$ be defined as in Algorithm~\ref{alg1:CSGM}. Then the MSE of $\hat{\mu}_\msf{CSGM}$ is bounded by
    $$ \msf{MSE}(\hat{\mu}_\msf{CSGM}) \eqDef \E\lb \lV \hat{\mu}_\msf{CSGM} - \mu \rV^2_2\rb \leq  \frac{\sigma^2}{n^2\gamma^2} + \frac{\Delta_2^2}{n\gamma}. $$
\end{corollary}
\looseness=-1
On the other hand, the MSE of the (uncompressed) Gaussian mechanism $\hat{\mu}_{\msf{GM}}$ is $ \msf{MSE}(\hat{\mu}_\msf{GM}) =  \sigma^2 / n^2$. It can be shown that under the same MSE constraints, the Renyi DP of $L_2$-CSGM converges to that of the Gaussian mechanism in the following sense: 
\begin{lemma}\label{lemma:compared_to_uncompressed_gaussian}
    For any fixed sparsification rate $\gamma$ and Renyi DP order $\alpha$,  let $\sigma^2_{\msf{GM}}$ and $\sigma^2_{\msf{CSGM}}$ be chosen such that $\msf{MSE}(\hat{\mu}_\msf{GM}) = \msf{MSE}(\hat{\mu}_\msf{CSGM})$, i.e., $\sigma^2_{\msf{GM}} = \frac{\sigma^2_{\msf{CSGM}}}{\gamma^2} + \frac{n\Delta^2_2}{\gamma}$. Then, it holds that $\varepsilon_{\msf{CSGM}}(\alpha) \ra \varepsilon_{\msf{GM}}(\alpha)$ as ${\Delta^2_\infty}/{\Delta^2_2} \ra 0$, where
    $\varepsilon_{\msf{GM}}(\alpha) = {\Delta_2^2\alpha} / {\sigma^2}$ is the R\'enyi DP bound of the Gaussian mechanism, and $\varepsilon_{\msf{CSGM}}(\alpha)$ is defined in \eqref{eq:l2_rdp_bdd}.
\end{lemma}

\begin{figure}[h]
\begin{center}
\centerline{\includegraphics[width=1.02\columnwidth]{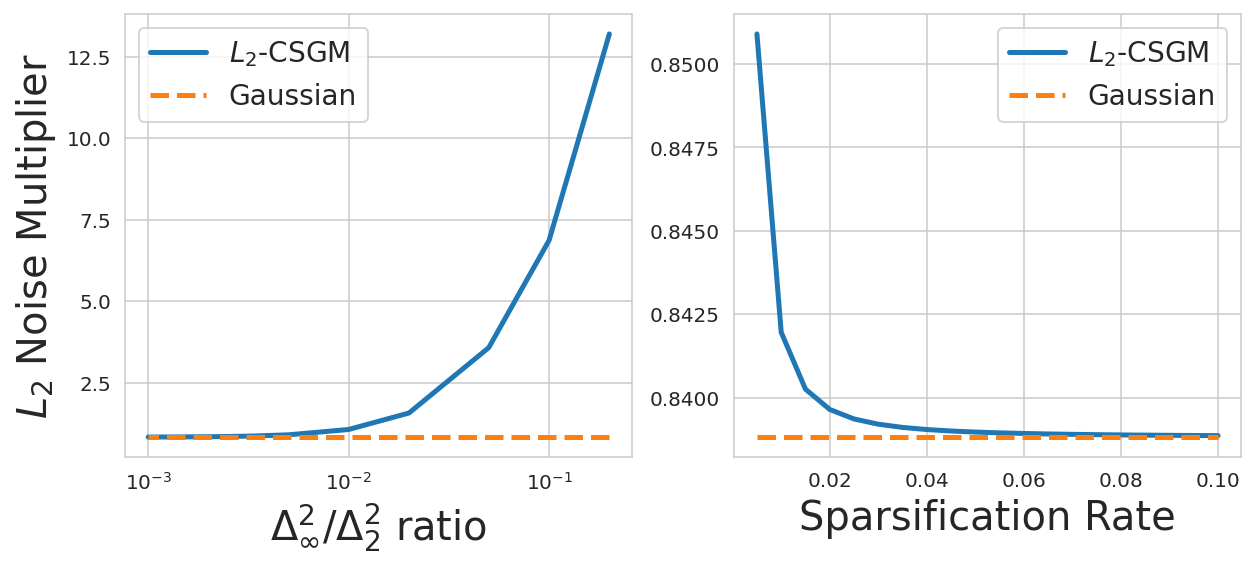}}
\vspace{-0.2in}
\caption{Noise multipliers (defined as $\sigma/\Delta_2$) of CSGM and GM with $\varepsilon = 5.0$, $\delta=10^{-8}$ and $\gamma=0.01$. On the left, we fix the sparsification rate $\gamma=0.01$. The numerical result indicates that as the ratio decreases, the noise multiplier of CSGM converges to that of the GM. Equivalently, this implies that $\varepsilon_{\msf{CSGM}}(\alpha) \ra \varepsilon_{\msf{GM}}(\alpha)$ if one fixes the MSEs of both schemes. On the right, we fix the $\Delta_2/\Delta_\infty$ ratio to be $1000$ and plot the noise multipliers.}
\label{fig:nmp_ratio}
\end{center}
\vspace{-0.2in}
\end{figure}

It is worth noting that, in general, the $\Delta_\infty/\Delta_2$ ratio decreases rapidly as $d$ increases, leading to $\varepsilon_{\msf{CSGM}}(\alpha) \rightarrow \varepsilon_{\msf{GM}}(\alpha)$ as $d \rightarrow \infty$. For instance, by utilizing random rotation for preprocessing local vectors, with high probability, $\Delta_\infty /\Delta_2 = O\left(\frac{\log d}{\sqrt{d}}\right)$. If further employing Kashin's representation \citep{lyubarskii2010uncertainty}, then $\Delta_\infty /\Delta_2 = O(1/\sqrt{d})$ with probability $1$.

\looseness=-1
On the other hand, if we calibrate the noise based on $\Delta_\infty$ as in $L_\infty$-CGSM, the constant in R\'enyi DP will not match that of the uncompressed Gaussian mechanism, which we elaborate on in the next subsection.

\looseness=-1
\subsection{Compared to $L_\infty$-CSGM \citep{chen2023privacy}.}\label{subsec:l_inf_comparison}
\looseness=-1
To compare the $L_2$ and $L_\infty$-CSGM, first observe that the R\'enyi DP bound in \eqref{eq:l2_rdp_bdd} can be expressed as $$ \varepsilon_{\msf{CGSM, L_2}}(\alpha) =  \Delta_2^2/\Delta_\infty^2\cdot D_\alpha\lp \Delta_\infty S  + Z \middle \Vert  Z  \rp, $$
where  $Z\sim \mcal{N}(0, \sigma^2)$ and  $S \sim \msf{Bern}(\gamma)$. On the other hand, the R\'enyi DP bound of $L_\infty$-CSGM in \citet{chen2023privacy} is $$ \varepsilon_{\msf{CGSM, L_\infty}}(\alpha) = d \cdot D_\alpha\lp \Delta_\infty S  + Z \middle \Vert  Z  \rp. $$
As a result, the ratio between two R\'enyi DP bounds is $ \frac{d\Delta^2_\infty}{\Delta^2_2} > 1$ (because $\lV \bm{g}\rV_\infty \leq \Delta_\infty$ implies $\lV \bm{g}\rV^2_2 \leq d\Delta^2_\infty$). When employing random rotation, this ratio is $O(\log(d))$ with high probability; with Kashin's representation, this ratio remains constant, but the constant is non-negligible (for instance, in \citet{chen2020breaking}, the constant is set to be around $2$). The sub-optimality gap between $L_\infty$-CSGM and the (uncompressed) Gaussian mechanism makes it undesirable in practical FL tasks, emphasizing the necessity of $L_2$-CSGM.

%% file: sec5_matrix_factorization.tex
\section{Matrix Factorization Mechanism with Local Sparsification under Streaming DP}
\looseness=-1
Moving on, we delve into the streaming DP setting, specifically focusing on the matrix mechanism detailed in Section~\ref{subsec:streaming_dp} and Section~\ref{subsec:matrix_mechanism}.

\looseness=-1
In the context of matrix mechanisms, the objective is to continually release a DP version of $\mb{A}\mb{G}$, where each row of $\mb{G}$ may depend on previous outputs $\bm{o}^{(1)},...,\bm{o}^{(t-1)}$. To minimize the overall MSE, $\lV \widehat{\mb{A}\mb{G}} - \mb{A}\mb{G} \rV_F^2$, we factorize $\mb{A}$ into $\mb{BC}$ and designing DP mechanisms according to $\mb{C}\mb{G}$, as discussed in Section~\ref{subsec:matrix_mechanism}. Notably, our scheme adopts the optimal factorization for the prefix sum matrix, addressing the optimization problem \eqref{eq:dp_factorization_problem}.

\looseness=-1
We aim to devise a matrix factorization scheme that simultaneously compresses local gradients $\mb{G}$. In this approach, instead of transmitting $\mb{G}$ to the server, clients send $\msf{compress}\lp\mb{G}\rp$, with compression applied row-wise (i.e., client-wise). A tempting strategy is to employ the local sparsification technique in CSGM and enhance privacy using Theorem~\ref{thm:mean_estimation_l2}\footnote{Throughout this section, we assume a cohort size of $1$ for simplicity. Our results naturally extend to general scenarios, and the full scheme is presented in Algorithm~\ref{alg:sgmf_full_cohort_size} in Appendix~\ref{appendix:sgmf_full_cohort_size}, in which each client adopts an independent sampling mask.}:
$ \mcal{M}_{\msf{SGMF}}\lp \mb{G} \rp \eqDef \mb{A}\lp \mb{S}\odot  \mb{G} \rp + \mb{BZ}, $
where $\mb{A} = \mb{BC}$ is a factorization, $[\mb{Z}]_{t, j} \diid \mcal{N}(0, \sigma^2)$ and $[\mb{S}]_{t, j} \diid \msf{Bern}(\gamma)$  for $t \in [T]$ and $j \in [d]$. However, the privacy analysis encounters two challenges: 
\looseness=-1
\begin{itemize}[leftmargin=1em,topsep=0pt]
    \setlength\itemsep{-0.5em}
    \looseness=-1
    \item In matrix mechanisms, a local vector $\bm{g}^{(t)}$ may persist across all $T$ rounds. Consequently, the randomness introduced in local sparsification steps at the $t$-th round might affect other rounds, resulting in what we term as temporal coupling. Unlike in \citet{denisov2022improved}, where the temporal coupling of isotropic Gaussian noise can be circumvented due to rotational invariance, local sparsification or sampling breaks this invariance, rendering Theorem~2.1 of \citet{denisov2022improved} inapplicable.

        \looseness=-1
    \item In the streaming scenario, the sampling variable $\bm{s}^{(t)}_j$ for the $j$-th coordinate in the $t$-round may influence the $j'$-th coordinate later due to adaptivity. For instance, $\bm{g}^{(t+1)}$ can depend on the $t$-th output $\bm{o}^{(t)}$, which, in turn, is a function of $\bm{s}^{(t)}_{j'}$ for all $j' \in [d]$. This introduces ``spatial correlation,'' which does not appear in the non-streaming setting (e.g., Theorem~\ref{thm:mean_estimation_l2}).
\end{itemize}

\vspace{-0.5em}
\begin{algorithm}
  \caption{Sparsified Gaussian Matrix Factorization}\label{alg:sgmf}
  \begin{algorithmic}[t]
  \State \textbf{Input:} Local vectors $\mb{g}^{(1)},...,\bm{g}^{(T)}$, noise scale $\sigma$, sparsification rate $\gamma$, factorization $\mb{A} = \mb{B}\cdot\mb{C}$.
  \State \texttt{//See Alg.~\ref{alg:sgmf_full_cohort_size} for cohort\_size > 1.}
      \For{Each client $t$ at time $t$}
        \State Generate $d$ independent binary masks $\mb{s}^{(t)} \in \{0, 1\}^d$:
        for any $j\in [d]$, $\mb{s}^{(t)}_j \diid \msf{Ber}(\gamma)$;
        \State Compute $\tilde{\mb{g}}^{(t)} = \bm{g}^{(t)}\odot \mb{s}^{(t)}$ and sends it to the server;
      \EndFor
    \State Server samples Gaussian noise: $[\mb{Z}]_{t, j} \diid \mcal{N}(0, \sigma^2)$ for all $t \in [T]$ and $j \in [d]$.
    \State Server computes the noisy outputs: $\mb{A} \cdot \tilde{\mb{G}} + \mb{B}\cdot\mb{Z}$.
  \end{algorithmic}
\end{algorithm}

\vspace{-0.5em}
In this section, we demonstrate that despite both temporal and spatial couplings, we can still achieve the same ``amplification effect'' as in Theorem~\ref{thm:mean_estimation_l2}. Our primary result is the R\'enyi Differential Privacy (DP) bound for the sparsified Gaussian matrix factorization outlined in Algorithm~\ref{alg:sgmf} (which can be seen as a direct extension of $L_2$-CSGM to the streaming DP setting).
\begin{theorem}\label{thm:sgmf_rdp}
Let $\mb{A} \in \mbb{R}^{T\times T}$ be a lower-triangular full-rank query matrix, and let $\mb{A} = \mb{BC}$ be any factorization for some $\mb{B}, \mb{C} \in \mbb{R}^{T\times T}$, with $\Delta(\mb{C}) = \max_{t \in [T]} \lV \bm{c}^{(t)} \rV_2$. Let $\mb{G}$ be the data matrix and $\Delta_2$ and $\Delta_\infty$ be the $L_2$ and $L_\infty$ norm bounds of $\mb{G}$, i.e., $\lV \bm{g}^{(t)}\rV_2 \leq \Delta_2$ and $\lV \bm{g}^{(t)} \rV_\infty \leq \Delta_\infty$ (recall that $\bm{g}^{(t)}$ denotes the $t$-th row of $\bm{G}$).
Then, the $\mcal{M}_{\msf{SGMF}}$ in Algorithm~\ref{alg:sgmf} satisfies adaptive $\lp \alpha, \varepsilon(\alpha) \rp$-R\'enyi DP for any $\alpha \geq 1$ and 
\begin{align}\label{eq:streaming_subsampled_dp_gaurantee}
        \varepsilon\lp \alpha \rp \geq &\frac{ \kappa_2^2/\kappa_\infty^2}{\alpha-1} \log\Big( (1-\gamma)^{\alpha-1}\lp \gamma(\alpha-1) + 1 \rp + \nonumber\\
        &\quad  \sum_{\ell=1}^\alpha { \alpha \choose \ell}\lp 1-\gamma  \rp^{\alpha - \ell} \gamma^{\ell} e^{(\ell-1) \ell\frac{\kappa_\infty^2}{2\sigma^2}} \Big).
\end{align}
where $\kappa_2 = \Delta(\mb{C})\cdot\Delta_2$ and $\kappa_\infty = \Delta(\mb{C})\cdot\Delta_\infty$ are the $L_2$ and $L_\infty$ sensitivities.
\end{theorem}

\looseness=-1
A couple of remarks follow. Firstly, the class of matrix mechanisms encompasses tree-based methods as a special case, such as $\msf{online}$ or $\msf{full}$-$\msf{honaker}$ tree aggregation \citep{honaker2015efficient} used in \citet{kairouz2021practical}. Therefore, Theorem~\ref{thm:sgmf_rdp} also applies to these results. Second, while \citet{choquette2023privacy} also investigate privacy amplification through subsampling, their subsampling is conducted client-wise rather than coordinate-wise, as their scheme does not aim for compression. Consequently, \citet{choquette2023privacy} do not encounter the spatial coupling issue. Finally, our scheme assumes single participation per epoch, and in practice, this can be done by shuffling and restarting the mechanism each epoch, similar to the TreeRestart approach in \citet{kairouz2021practical}.

\looseness=-1
\subsection{Proof of Theorem~\ref{thm:sgmf_rdp}}
\looseness=-1
Next, we prove Theorem~\ref{thm:sgmf_rdp}. The proof begins with the LQ decomposition trick in \citet{denisov2022improved}, followed by a careful decoupling of the joint distribution on $\bm{o}^{(1)}, ..., \bm{o}^{(t)}$.

\looseness=-1
\paragraph{Reparameterization.} Let $\mb{B} = \mb{L}\cdot\mb{Q}$ be the LQ decomposition of the matrix $\mb{B}$. Consider a different lower-triangular factorization:
$ \tilde{\mcal{M}}\lp \mb{G}\rp = \mb{L}\lp \mb{QC}\lp\mb{G}\odot\mb{S}\rp + \mb{Z} \rp, $
where $\mb{Z}_{ij} \diid \mcal{N}(0, \sigma^2)$, $\mb{Q}$ is orthonormal, and both $\mb{L}$ and $\mb{QC}$ are lower-triangular. Since $\mb{QC}$ is lower-triangular,  $\mb{QC}\lp\mb{G}\odot\mb{S}\rp + \mb{Z}$ can operate in the continuous release model, as row $t$ of $\mb{QC}(\mb{G}\odot\mb{S})$ depends only on the first $t$ rows of $\mb{G}$. Following from the same argument in \citet[Theorem~2.1]{denisov2022improved}, it suffices to show the desired DP guarantee \eqref{eq:streaming_subsampled_dp_gaurantee} on $\mb{QC}\lp\mb{G}\odot\mb{S}\rp + \mb{Z}$ since we can always replace $\mb{Z}$ with $\mb{QZ}$ due to the rotational invariance of isotropic Gaussian distribution. For notational convenience, we denote $\mb{QC} \eqDef \mb{M}$ in the remaining proof (note that $\mb{M}$ is lower triangular). 

\looseness=-1
\paragraph{Joint density of the transcript.} Next, we show the mechanism $\mb{M}\lp\mb{G}\odot\mb{S}\rp + \mb{Z}$ is an instance of the standard (subsampled) Gaussian mechanism for computing an adaptive function in the continuous release model with a guaranteed bound on the global $L_2$ and $L_\infty$ sensitivities. Let $\mb{G}$ and $\mb{H}$ be any two neighboring data streams (defined in Definition~\ref{def:streaming_neighboring}) that additionally satisfy the following $L_\infty$ condition: $\max_{t \in [T], \, j \in [d]} \lba \mb{g}^{(t)}_j - \mb{h}^{(t)}_j\rba \leq \Delta_\infty$.
 Without loss of generality, we assume that $\mb{G}$ and $\mb{H}$ differ at $t = 1$, and thus when analyzing the privacy guarantees, we condition on the realization $(\mb{s}^{(2)},...,\mb{s}^{(T)}) = (\check{\mb{s}}^{(2)}, ..., \check{\mb{s}}^{(T)})$ and all the potential randomness used in the optimization algorithm, treating them as deterministic. The only randomness that will be accounted for in the privacy analysis is $\mb{s}^{(1)}$ and $\mb{Z}$.

Given the data stream $\mb{G}$, the output transcript $\mb{O} \eqDef \lp \bm{o}^{(1)}, \bm{o}^{(2)}, ..., \bm{o}^{(T)} \rp \in \mbb{R}^{d\times T}$ is computed as follows:

\begin{align*}
    &\bm{o}^{(1)} = \underbrace{\mb{M}_{11}\lp \bm{g}^{(1)}\odot\mb{s}^{(1)}\rp + \mb{Z}^{(1)}}_{\eqDef \mb{p}^{(1)}};\\
    & \bm{o}^{(t)} = \underbrace{\mb{M}_{t1} \lp \bm{g}^{(1)}\odot\mb{s}^{(1)}\rp + \mb{Z}^{(t)}}_{\eqDef \bm{p}^{(t)}} + \underbrace{\sum_{\tau=2}^{t} \mb{M}_{t\tau}\lp \bm{g}^{(\tau)} \odot \check{\mb{s}}^{(\tau)} \rp}_{\eqDef \bm{q}^{(t)}},
\end{align*}

for all $t \geq 1$. Our goal is to control
$ D_{\alpha}\lp P_{\mb{O}|\mb{G}} \middle \Vert P_{\mb{O}|\mb{H}} \rp, $
where $P_{\mb{O}|\mb{G}}$ denotes the distribution of transcript $\mb{O}$ under data stream $\mb{G}$ and $P_{\mb{O}|\mb{H}}$ denotes the distribution of $\mb{O}$ under $\mb{H}$. Note that the randomness used to compute the above divergence only includes $\mb{Z}^{(1)}, ..., \mb{Z}^{(T)}$ and $\mb{s}^{(1)}$, as we have conditioned on all other (irrelevant) external randomness, including $\check{\mb{s}}^{(2)}, ..., \check{\mb{s}}^{(T)}$.

\paragraph{Decoupling the joint distribution.} The main challenge here, compared to the uncompressed Gaussian mechanism in \citet{denisov2022improved}, is the spatial and temporal coupling on the joint distribution $P_{\mb{O}|\mb{G}}$. To see this, observe that $\bm{o}^{(t)}_{j'}$ implicitly depends on the $j$-th sampling variable $\mb{s}^{(1)}_j$ through $\bm{g}^{(2)}, ..., \bm{g}^{(t-1)}$ (which are functions of $\bm{o}^{(1)}, ..., \bm{o}^{(t-1)}$). As a result, the joint distribution of $\mb{O}$ is a mixture of product distributions, so the scheme cannot be reduced into a simple subsampled Gaussian mechanism.

To address this issue, we introduce the following decomposition trick on the transcript $\bm{o}^{(t)}$ to decouple the complicated spatial and temporal correlation. For all $t \geq 1$, write $\bm{p}^{(t)} \eqDef \mb{M}_{t1} \lp \bm{g}^{(1)}\odot\mb{s}^{(1)}\rp + \mb{Z}^{(t)}$,  $\bm{q}^{(1)} \eqDef 0$, and
$$   \bm{q}^{(t)} \eqDef \sum_{\tau=2}^{t} \mb{M}_{t\tau}\lp \bm{g}^{(\tau)} \odot \check{\mb{s}}^{(\tau)} \rp, $$
so that $\bm{o}^{(t)} = \bm{p}^{(t)} + \bm{q}^{(t)}$.

The key observation is that, conditioned on the realization $\check{\mb{s}}^{(2)}, ..., \check{\mb{s}}^{(T)}$, $\mb{Q} \eqDef (\bm{q}^{(1)}, ..., \bm{q}^{(T)})$ is a \emph{deterministic} function of $\mb{P} \eqDef (\bm{p}^{(1)}, ..., \bm{p}^{(T)})$. To see this, note that
\begin{align*}
    \bm{q}^{(t)}  = f(\bm{o}^{(1)}, ..., \bm{o}^{(t-1)})
    = g\lp (\bm{p}^{(1)}, \bm{q}^{(1)}), ..., (\bm{p}^{(t-1)}, \bm{q}^{(t-1)})\rp
\end{align*}
for some functions $f$ and $g$. Also notice that $\bm{q}^{(1)} = 0$. Thus, by induction, $\bm{q}^{(t)}$ is a function of $\bm{p}^{(1)}, ..., \bm{p}^{(t-1)}$.

As a result, the overall transcript $\mb{O} = \mb{P} + \mb{Q}(\mb{P})$ can be viewed as a post-processing of $\mb{P}$, so by data processing inequality, it holds that
\begin{align}\label{eq:decoupling_p}
D_{\alpha}\lp P_{\mb{O}|\mb{G}} \middle \Vert P_{\mb{O}|\mb{H}} \rp \leq D_{\alpha}\lp P_{\mb{P}|\mb{G}} \middle \Vert P_{\mb{P}|\mb{H}} \rp.
\end{align}
Since the $\mb{P} = (\bm{p}^{(1)}, ..., \bm{p}^{(T)})$ does not have spatial coupling, in the sense that $\bm{p}^{(t)}_j$ is independent of $\mb{s}^{(1)}_{j'}$ for all $t \in [T]$ and $j, j' \in [d]$, $j \neq j'$, we can invoke the argument of \citet{denisov2022improved} along with privacy amplification by subsampling, summarized as in the following lemma.
\begin{lemma}\label{lemma:PG_PH_rdp_bound}
    Let $\mb{P}$ be defined as above. Then, it holds that
    \begin{align*}
    D_{\alpha}\lp P_{\mb{P}|\mb{G}} \middle \Vert P_{\mb{P}|\mb{H}} \rp \leq \frac{ \kappa_2^2/\kappa_\infty^2}{\alpha-1} \log\Big( (1-\gamma)^{\alpha-1}\cdot
 \lp \gamma(\alpha-1) + 1 \rp + \sum_{\ell=1}^\alpha { \alpha \choose \ell}\lp 1-\gamma  \rp^{\alpha - \ell} \gamma^{\ell} e^{(\ell-1) \ell\frac{\kappa_\infty^2}{2\sigma^2}} \Big).
\end{align*}
\end{lemma}

\looseness=-1
We remark that \eqref{eq:decoupling_p} implies that among all possible adaptive dependencies of $\bm{g}^{(t)}(\bm{o}^{(1)}, ..., \bm{o}^{(t-1)})$, the transcript $\mb{O}$ is statistically dominated by the independent one, that is, $\bm{g}^{(t)}$ remains constant regardless of previous outputs $(\bm{o}^{(1)}, ..., \bm{o}^{(t-1)})$. 
\qedwhite

%% file: sec6_experiments.tex
\section{Empirical Evaluation}
We provide empirical evaluations on the privacy-utility trade-offs for both DP-SGD (under a non-streaming setting) and DP-FTRL type (with matrix mechanisms \cite{denisov2022improved}) algorithms. We mainly compare the $L_2$-CGSM (Algorithm~\ref{alg:sgmf}) and sparsified Gaussian matrix factorization (Algorithm~\ref{alg:sgmf}) with the uncompressed Gaussian mechanism \citep{balle2018improving}. We convert the R\'enyi DP bounds to $(\varepsilon, \delta)$-DP via the conversion lemma from \citet{canonne2020discrete} for a fair comparison.

\textbf{Datasets and models.}  We run experiments on the full Federated EMNIST~\citep{cohen2017emnist} and Stack Overflow~\citep{stackoverflow2019} dataset. F-EMNIST has $62$ classes and $N=3400$ clients with a total of $671,585$ training samples. Inputs are single-channel $(28,28)$ images. The Stack Overflow (SO) dataset is a large-scale text dataset based on responses to questions asked on the site Stack Overflow. There are over $10^8$ data samples unevenly distributed across $N=342,477$ clients. We focus on the next word prediction (NWP) task: given a sequence of words, predict the next words in the sequence.

On F-EMNIST, we experiment with a (4 layer) Convolutional Neural Network (CNN) used by~\citet{kairouz2021distributed} (with around $1$ million parameters). On SONWP, we experiment with a $4$ million parameters ($4$ layer) long-short term memory (LSTM) model -- the same as prior work~\citet{andrew2021differentially,kairouz2021distributed}.
In both cases, clients train for $1$ local epoch using SGD. Only the server uses momentum.

Additionally, for local model updates, we perform random rotation and $L_\infty$-clipping, with $\Delta_\infty = \Delta_2\sqrt{2\log(d\cdot n)/d}$, where $d$ is the model dimension (i.e., $\#$ trainable parameters) and $n$ is the cohort size in each training round.

\begin{figure}[!th]
\begin{center}
\centerline{\includegraphics[width=0.75\columnwidth]{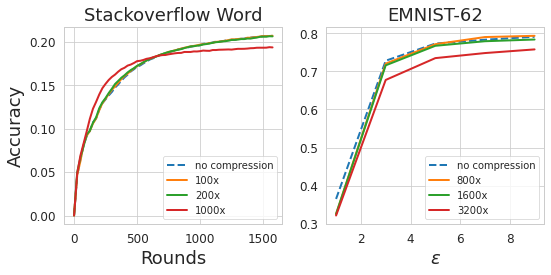}\vspace{-0.18in}}
\caption{Accuracy of GM and CSGM, with $\delta = 10^{-5}$ for F-EMNIST and $\delta = 10^{-6}$ for SONWP. The resulting $\Delta_\infty/\Delta_2$ value is $6.4\cdot 10^{-3}$ for F-EMNIST and $3.3\cdot 10^{-3}$ for SONWP.}
\label{fig:csgm}
\end{center}
\vskip -0.25in
\end{figure}

\paragraph{$L_2$-CSGM for DP-SGD.}
In Figure~\ref{fig:csgm}, we report the accuracy of $L_2$-CSGM (Algorithm~\ref{alg1:CSGM}) as well as the uncompressed Gaussian mechanism.

\begin{figure}[!th]
\begin{center}
\centerline{\includegraphics[width=0.75\columnwidth]{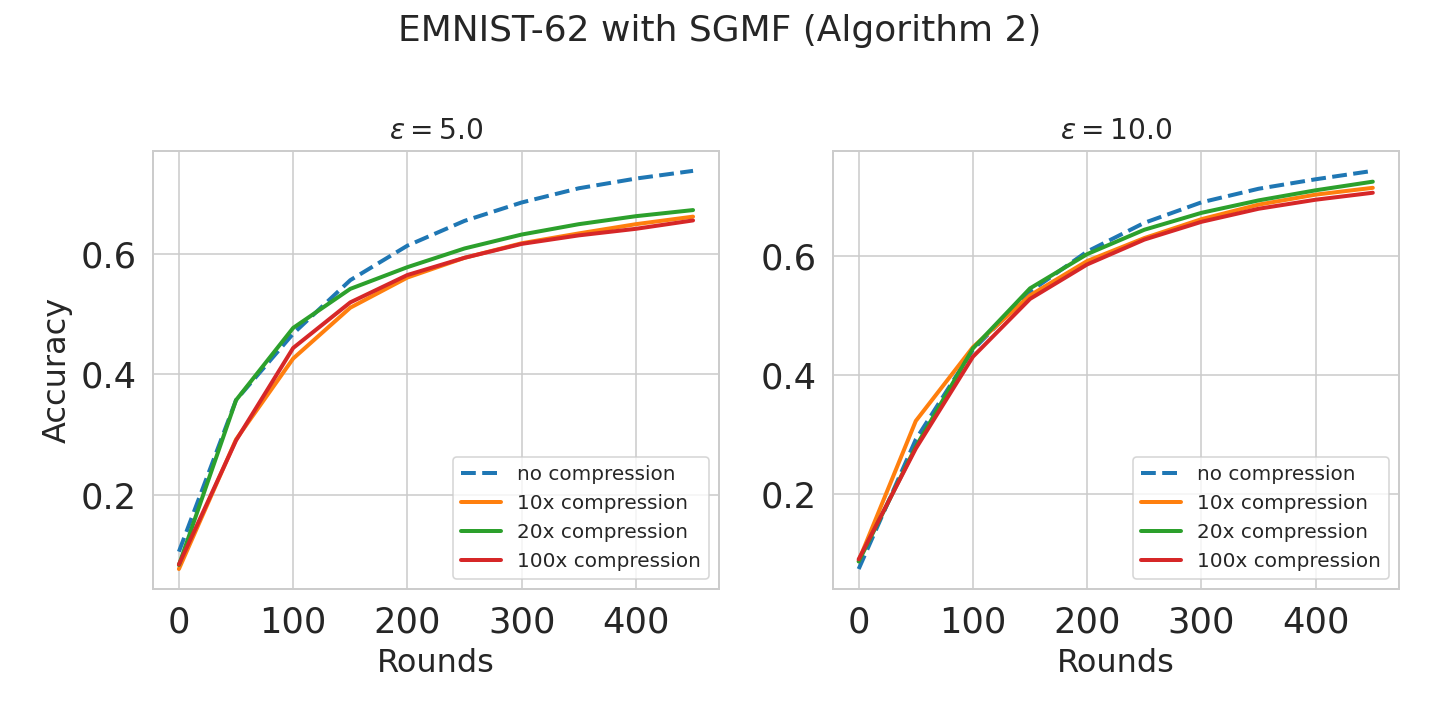}}\vspace{-0.18in}
\caption{Accuracy of MF and SGMF, with $\delta = 10^{-5}$, cohort size $n= 100$, clipped norm $\Delta_2 = 1.0$, and server learning rate $0.1$.}
\label{fig:sgmf}
\end{center}
\vskip -0.25in
\end{figure}

\textbf{Sparsified Gaussian Matrix Mechanism for DP-FTRL.}
In Figure~\ref{fig:sgmf}, we report the accuracy of SGMF (Algorithm~\ref{alg1:CSGM}) and the uncompressed matrix mechanism. We use the same optimal factorization as in \citet{denisov2022improved} with $T=32$ for 16 epochs, and we restart the mechanism and shuffling clients every epoch as in the $\msf{TreeRestart}$ approach in \citet{kairouz2021practical}. We observe that for the matrix mechanism, the compression rates are, in general, less than DP-FedAvg, and in addition, the performance is more sensitive to server learning rates and $L_2$ clip norms.

%% file: sec7_conclusion.tex
\vspace{-1.5em}
\section{Conclusion}
Our work addresses challenges in $L_2$ mean estimation under central DP and communication constraints. We introduce a novel $L_2$ R\'enyi DP accounting algorithm for the sparsified Gaussian mechanism that significantly improves upon previous ones based on $L_\infty$ sensitivity. We also extend the scheme and accountant to the streaming setting, providing an adaptive DP bound that handles spatial and temporal couplings of privacy loss unique to adaptive settings. Empirical evaluations on diverse federated learning tasks showcase a 100x enhancement in compression. Notably, our scheme focuses on reducing the dimensionality of local model updates, and hence it can potentially be combined with other gradient quantization or compression techniques, thereby promising heightened compression efficiency.

%% file: sec99_appendix.tex
\appendix

\section{Sparsified Gaussian Matrix Factorization for General Cohort Size}\label{appendix:sgmf_full_cohort_size}
In this section, we present the full SGMF schemes with a general cohort size. Note that while we allow more than one client per FL round, each client only participates \emph{once}.
\begin{algorithm}
  \caption{Sparsified Gaussian Matrix Factorization with Full Cohort Size}\label{alg:sgmf_full_cohort_size}
  \begin{algorithmic}[t]
  \State \textbf{Input:} Local vectors $\mb{g}^{(1)},...,\bm{g}^{(T)}$, noise scale $\sigma$, sparsification rate $\gamma$, factorization $\mb{A} = \mb{B}\cdot\mb{C}$.
      \For{Each {\color{blue}cohort $\mcal{B}_t$} at time $t$}
      \For{Each {\color{blue} client $i$ in cohort $\mcal{B}_t$}}
        \State Generates an independent binary mask $\mb{s}^{(t, i)} \in \msf{Ber}(\gamma)^{\otimes d}$;
        Send $\tilde{\mb{g}}^{(t, i)} = \bm{g}^{(t, i)}\odot \mb{s}^{(t, i)}$ to the server;
      \EndFor
      \EndFor
    \State Server computes $\tilde{G} \in \mbb{R}^{T\times d}$, where the $t$-th row is $\tilde{\mb{g}}^{(t)} = \frac{1}{\gamma |\mcal{B}_t|}\sum_{i \in \mcal{B}_t}\tilde{\mb{g}}^{(t, i)}$;
    \State Server samples Gaussian noise: $[\mb{Z}]_{t, j} \diid \mcal{N}(0, \sigma^2)$ for all $t \in [T]$ and $j \in [d]$.
    \State Server computes the noisy mean: $\mb{A} \cdot \tilde{\mb{G}} + \mb{B}\cdot\mb{Z}$.
  \end{algorithmic}
\end{algorithm}

\section{Additional Details on Communication-Efficient DME with Local DP}\label{sec:for_apple_folks}
An alternative method for achieving communication-efficient DME under central DP involves employing local DP mechanisms \citep{warner1965randomized, kasiviswanathan2011can} with privacy amplification through shuffling \citep{erlingsson2019amplification, girgis2021shuffled, feldman2022hiding, feldman2023stronger}. It is worth noting that, under a $\varepsilon_\msf{Local}$-DP constraint, the optimal local randomizer is the privUnit mechanism \citep{bhowmick2018protection, asi2022optimal}. This mechanism can be efficiently compressed using a pseudo-random generator (PRG) \citep{feldman2021lossless} or random projection \citep{asi2023fast} (without going through quantization or $L_\infty$ clipping). Combining these local DP schemes with a multi-message shuffler has been proven to achieve order-optimal privacy-accuracy-utility trade-offs \citep{chen2023privacy, girgis2023multi}, requiring less trust assumption on the server.

However, as pointed out in \citet{chen2023privacy}, this local DP approach involves privacy amplification by shuffling lemmas that exhibit large leading constants compared to CGSM. Furthermore, privUnit is designed and optimized under \emph{pure} DP, leaving its optimality under approximate or R\'enyi DP unclear. Additionally, to our best knowledge, there is currently no privacy amplification lemma known for transforming local R\'enyi DP into central R\'enyi DP. Hence, even if one adopts an optimal \emph{local R'enyi} DP scheme and combines it with shuffling, it remains uncclear whether the resulting privacy guarantee is order-optimal. Lastly, \citet{chen2023privacy} empirically demonstrates a non-negligible gap in Mean Squared Errors (MSEs) between shuffling-based methods and $L_\infty$-CGSM.

\section{Additional Details for the Experiments}
In this section, we provide additional details of the experiments. We mainly compare the $L_2$-CGSM (Algorithm~\ref{alg:sgmf}) and sparsified Gaussian matrix factorization (Algorithm~\ref{alg:sgmf}) with the uncompressed Gaussian mechanism \citep{balle2018improving}. We convert the R\'enyi DP bounds to $(\varepsilon, \delta)$-DP via the conversion lemma from \citet{canonne2020discrete} for a fair comparison.

\textbf{Datasets and models.}  We run experiments on the full Federated EMNIST~\citep{cohen2017emnist} and Stack Overflow~\citep{stackoverflow2019} dataset. F-EMNIST has $62$ classes and $N=3400$ clients with a total of $671,585$ training samples. Inputs are single-channel $(28,28)$ images. The Stack Overflow (SO) dataset is a large-scale text dataset based on responses to questions asked on the site Stack Overflow. There are over $10^8$ data samples unevenly distributed across $N=342,477$ clients. We focus on the next word prediction (NWP) task: given a sequence of words, predict the next words in the sequence.

On F-EMNIST, we experiment with a (4 layer) Convolutional Neural Network (CNN), which is used by~\citet{kairouz2021distributed}. The architecture is slightly smaller and has $d \leq 2^{20}$ parameters to reduce the zero padding required by the randomized Hadamard transform used for flattening and $L_\infty$ clipping (see Algorithm~\ref{alg1:CSGM}). The requirement can be potentially removed if one uses a randomized Fourier transform instead. On SONWP, we experiment with a $4$ million parameters ($4$ layer) long-short term memory (LSTM) model -- the same architecture as prior work \citet{andrew2021differentially,kairouz2021distributed}.
In both cases, clients train for $1$ local epoch using SGD. Only the server uses momentum.

For each local model update, we perform random rotation (based on randomized Hadamard transform) and $L_\infty$ clipping, with $\Delta_\infty = \Delta_2\sqrt{2\log(d\cdot n)/d}$, where $d$ is the model dimension (i.e., $\#$ trainable parameters) and $n$ is the cohort size in each training round.

\paragraph{$L_2$-CSGM for DP-SGD.}

\begin{figure}[th]
    \centering
    \begin{minipage}[t]{0.5\linewidth}
    \includegraphics[width=\linewidth]{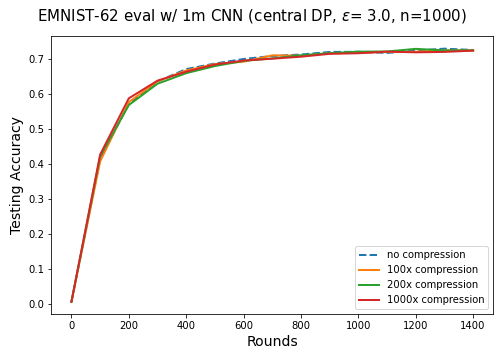}
    \end{minipage}\hfill
    \begin{minipage}[t]{0.5\linewidth}
    \includegraphics[width=\linewidth]{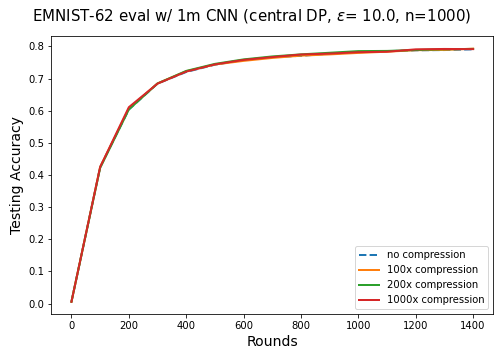}
    \end{minipage}
    \caption{Accuracy of GM and CSGM, with $\delta = 10^{-5}$ and cohort size $1000$. The $\Delta_\infty/\Delta_2$ ratio is $6.4\cdot 10^{-3}$ for F-EMNIST.}\label{fig:csgm_emnist_1}
\end{figure}

\begin{figure}[th]
    \centering
    \begin{minipage}[t]{0.5\linewidth}
    \includegraphics[width=\linewidth]{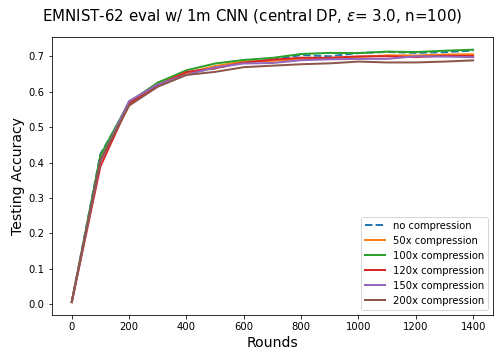}
    \end{minipage}\hfill
    \begin{minipage}[t]{0.5\linewidth}
    \includegraphics[width=\linewidth]{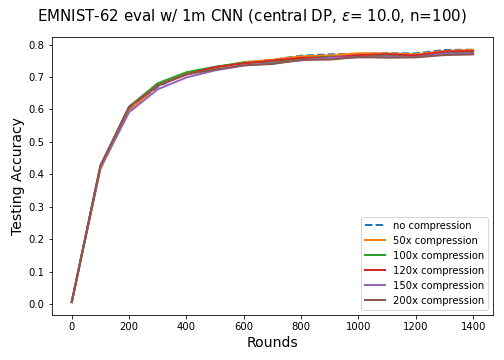}
    \end{minipage}
    \caption{Accuracy of GM and CSGM, with $\delta = 10^{-5}$ and cohort size $100$. The $\Delta_\infty/\Delta_2$ ratio is $6.4\cdot 10^{-3}$ for F-EMNIST.}\label{fig:csgm_emnist_2}
\end{figure}

\begin{figure}[th]
    \centering
    \begin{minipage}[t]{0.45\linewidth}
    \includegraphics[width=\linewidth]{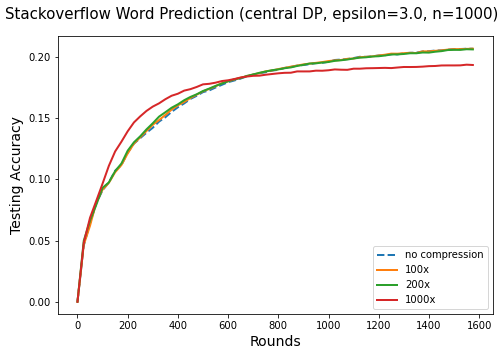}
    \end{minipage}\hfill
    \begin{minipage}[t]{0.45\linewidth}
    \includegraphics[width=\linewidth]{figures/fedavg/sonwp_eps_3.0_n_1000}
    \end{minipage}
    \caption{Accuracy of GM and CSGM, with $\delta = 10^{-5}$ and cohort size $100$. The $\Delta_\infty/\Delta_2$ ratio is $6.4\cdot 10^{-3}$ for F-EMNIST.}\label{fig:csgm_sonwp}
\end{figure}

In Figure~\ref{fig:csgm_emnist_1} and Figure~\ref{fig:csgm_emnist_2}, we present the accuracy results of the $L_2$-CSGM algorithm (Algorithm \ref{alg1:CSGM}) applied to F-EMNIST with varying cohort sizes, juxtaposed with the performance of the uncompressed Gaussian mechanism. Notably, our findings reveal that, on the whole, we can achieve compression exceeding 100x without a significant compromise in accuracy. Furthermore, as the cohort size $n$ increases, the impact of compression on utility diminishes. This implies that greater compression is feasible with larger values of $n$. Similarly, in Figure \ref{fig:csgm_sonwp}, we delineate the accuracy outcomes for the Stack Overflow next-word prediction task across diverse $\varepsilon$ values, maintaining a constant cohort size of $1000$.

\begin{figure}[th]
    \centering
    \begin{minipage}[t]{0.45\linewidth}
    \includegraphics[width=\linewidth]{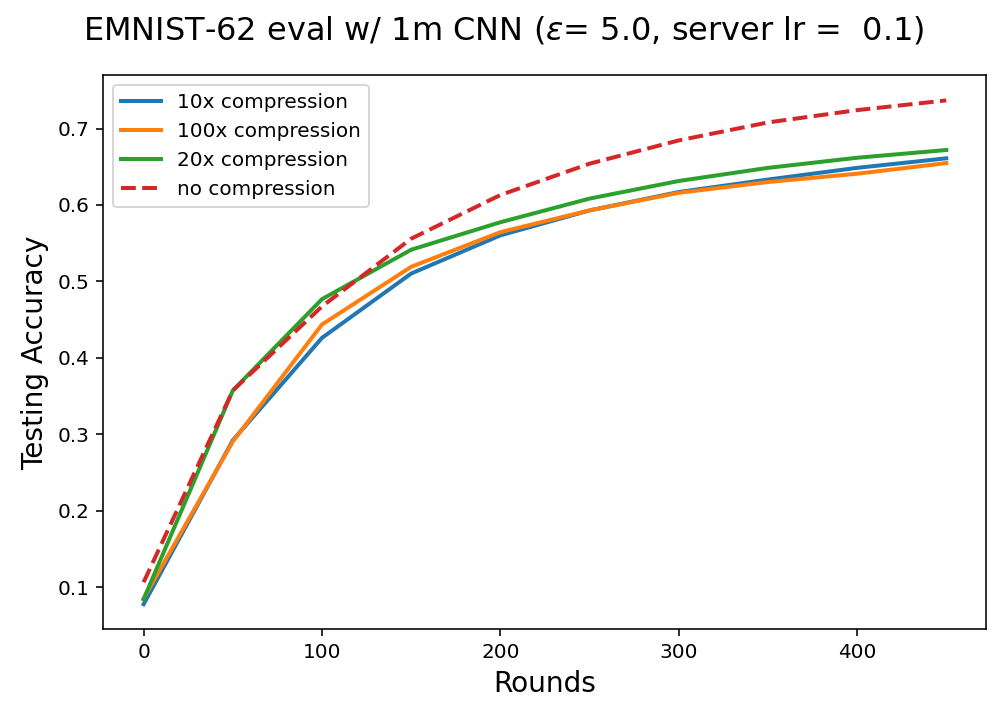}
    \end{minipage}\hfill
    \begin{minipage}[t]{0.45\linewidth}
    \includegraphics[width=\linewidth]{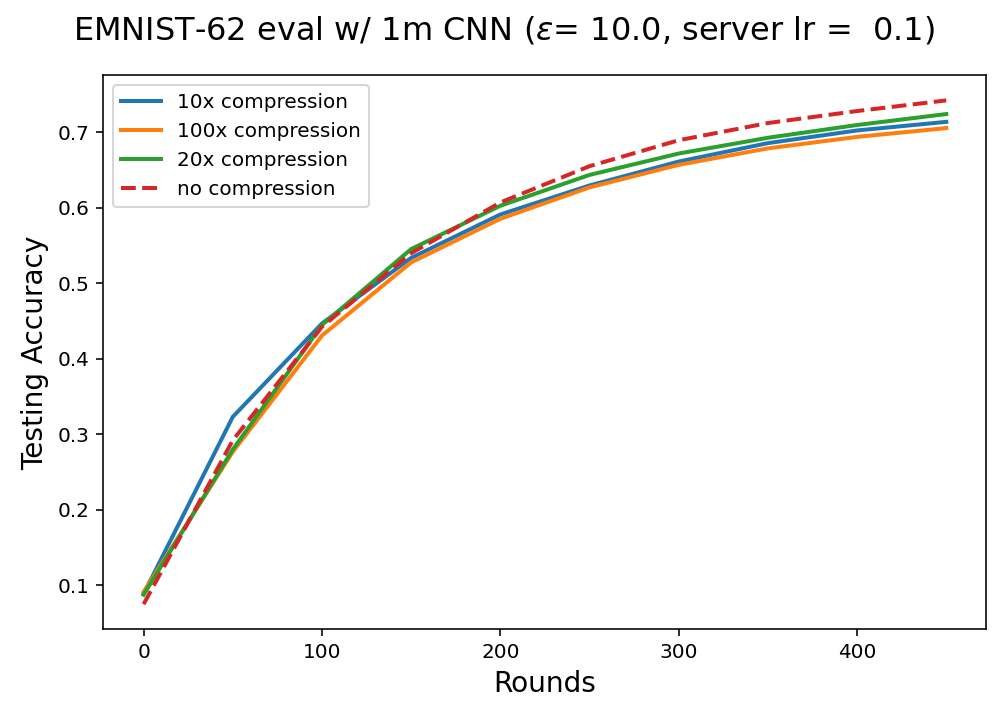}
    \end{minipage}
    \caption{Accuracy of MF and SGMF on EMNIST, with $\delta = 10^{-5}$,  clipped norm $\Delta_2 = 1.0$, and server learning rate $0.1$.}\label{fig:sgmf_1}
\end{figure}

\begin{figure}[ht]
    \centering
    \begin{minipage}[t]{0.45\linewidth}
    \includegraphics[width=\linewidth]{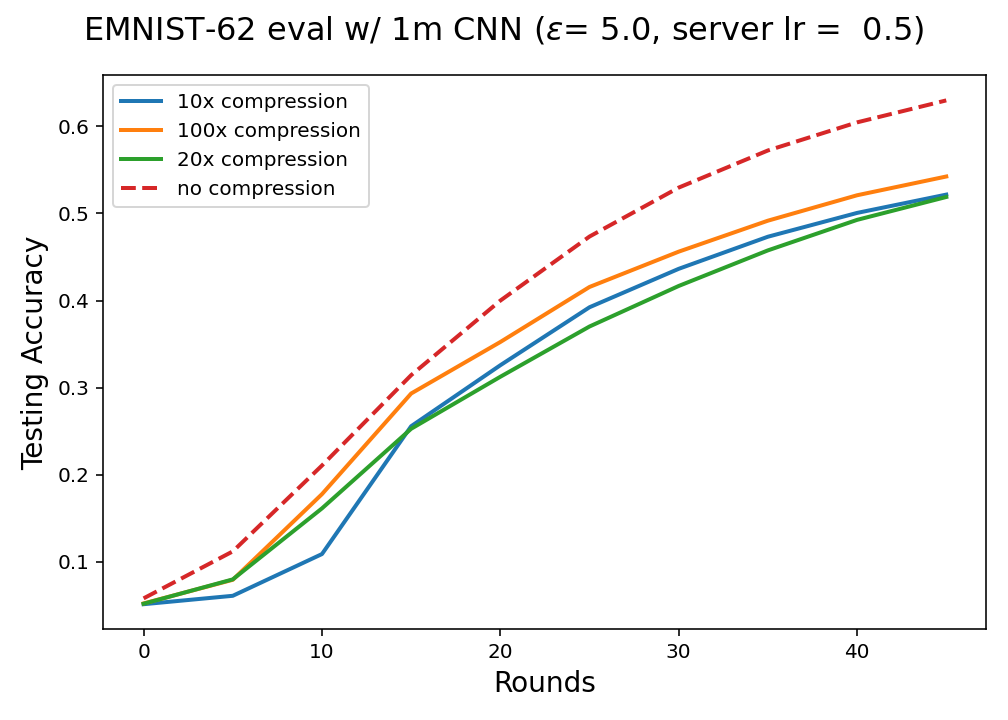}
    \end{minipage}\hfill
    \begin{minipage}[t]{0.45\linewidth}
    \includegraphics[width=\linewidth]{figures/mf/emnist_full_rounds_512_eps_5.0_sparsification_1.0_slr_0.5_clr_0.32_clip_0.03_mf}
    \end{minipage}
    \caption{Accuracy of MF and SGMF on EMNIST, with $\delta = 10^{-5}$,  clipped norm $\Delta_2 = 0.03$, and server learning rate $0.5$.}\label{fig:sgmf_2}
\end{figure}

\textbf{Sparsified Gaussian Matrix Mechanism for DP-FTRL.}
In Figure~\ref{fig:sgmf_1} and Figure~\ref{fig:sgmf_2}, we report the accuracy of SGMF (Algorithm~\ref{alg:sgmf}) and the uncompressed matrix mechanism. We use the same factorization as in \citet{denisov2022improved} with $T=32$ for $16$ epochs (due to the limited amount of clients), and we restart the mechanism and shuffle clients every epoch as in the $\msf{TreeRestart}$ approach in \citet{kairouz2021practical}. We observe that for the matrix mechanism, the compression rates are generally less than DP-FedAvg, and the performance is more sensitive to server learning rates and $L_2$ clip norms.

\section{Proofs}
\subsection{Proof of Theorem~\ref{thm:mean_estimation_l2}}
For any $\bm{g}_1, \bm{g}_2, ..., \bm{g}_n$, it holds that
\begin{align*}
    &D_{\alpha}\lp \bm{s}_1\odot \bm{g}_1 + \sum_{i=2}^n \bm{s}_i\odot \bm{g}_i + Z \middle\Vert  \sum_{i=2}^n \bm{s}_i\odot \bm{g}_i + Z \rp\\
    & \overset{\text{(a)}}{\leq} D_{\alpha}\lp  \bm{s}_1\odot \bm{g}_1  + Z \middle\Vert  Z \rp\\
    & \overset{\text{(b)}}{=}  \sum_{j=1}^d D_{\alpha}\lp  \bm{s}_1(j) \cdot \bm{g}_{1}(j)  + Z_j \middle\Vert   Z_j \rp\\
    & = \sum_{j=1}^d D_{\alpha}\lp  \gamma\mcal{N}(\bm{g}_{1}(j), \sigma^2)  + (1-\gamma) \mcal{N}(0, \sigma^2)  \middle\Vert  \mcal{N}(0, \sigma^2) \rp,
\end{align*}
where (a) is due to the data processing inequality, 
and in (b) holds since $ \bm{s}_1(j)$ and $Z_j$ are independent across $j \in [d]$. 
Similarly, it holds that
\begin{align*}
    &D_{\alpha}\lp \sum_{i=2}^n \bm{s}_i\odot \bm{g}_i + Z \middle\Vert \bm{g}_1 + \sum_{i=2}^n \bm{s}_i\odot\bm{g}_i + Z \rp \leq \sum_{j=1}^d D_{\alpha}\lp  \mcal{N}(0, \sigma^2)  \middle\Vert  \gamma\mcal{N}(\bm{g}_{1}(j), \sigma^2)  + (1-\gamma) \mcal{N}(0, \sigma^2)  \rp,
\end{align*}
For notational simplicity, let us define $\kappa_j \eqDef\bm{g}_{1}(j)$. Notice that $\bm{g}_1\in \mbb{S}^{d-1}$ implies $\lV \kappa \rV_2 \leq 1$.

Then for each $j \in [d]$, by Corollary~7 of \citet{mironov2019r},
\begin{align*}
    &\max\lp D_{\alpha}\lp  \gamma\mcal{N}(x_{1j}, \sigma^2)  + (1-\gamma) \mcal{N}(0, \sigma^2)  \middle\Vert  \mcal{N}(0, \sigma^2) \rp,   D_{\alpha}\lp  \mcal{N}(0, \sigma^2)  \middle\Vert  \gamma\mcal{N}(x_{1j}, \sigma^2)  + (1-\gamma) \mcal{N}(0, \sigma^2)  \rp \rp\\
     &= D_{\alpha}\lp  \gamma\mcal{N}(x_{1j}, \sigma^2)  + (1-\gamma) \mcal{N}(0, \sigma^2)  \middle\Vert  \mcal{N}(0, \sigma^2) \rp\\
     &=  \frac{1}{\alpha-1}\log\lp \E_{X\sim q}\lb \lp (1-\gamma) + \gamma\frac{p}{q}(X) \rp^\alpha\rb\rp 
\end{align*}
where $p$ is a density function of $\mcal{N}(\kappa_j, \sigma^2)$ and $q$ is a density function of $\mcal{N}(0, \sigma^2)$.

For any integer $\alpha$, we have
\begin{align*}
     & \frac{1}{\alpha-1}\log\lp \E_{X\sim q}\lb \lp (1-\gamma) + \gamma\frac{p}{q}(X) \rp^\alpha\rb\rp \\
     & = \frac{1}{\alpha -1} \log\lp 
     \sum_{\ell=0}^\alpha { \alpha \choose \ell}\lp 1-\gamma  \rp^{\alpha - \ell} \gamma^{\ell} 
     \E_{X\sim q}\lb\exp\lp\ell\lp -\frac{1}{2\sigma^2} \rp ((X-\kappa_j)^2 - X^2) \rp\rb\rp\\
     & = \frac{1}{\alpha -1} \log\lp 
     \sum_{\ell=0}^\alpha { \alpha \choose \ell}\lp 1-\gamma  \rp^{\alpha - \ell} \gamma^{\ell} 
     \E_{X\sim q}\lb\exp\lp -\frac{ \ell}{2\sigma^2} (\kappa_j^2 - 2\kappa_j X) \rp\rb\rp\\
     & = \frac{1}{\alpha -1} \log\lp 
     \sum_{\ell=0}^\alpha { \alpha \choose \ell}\lp 1-\gamma  \rp^{\alpha - \ell} \gamma^{\ell} 
     \exp\lp\frac{-\ell \kappa_j^2}{2\sigma^2}\rp
     \E_{X\sim q}\lb\exp\lp \frac{\kappa_j \ell}{\sigma^2}  X \rp\rb\rp\\
     & \overset{\text{(a)}}{=} \frac{1}{\alpha -1} \log\lp 
     \sum_{\ell=0}^\alpha { \alpha \choose \ell}\lp 1-\gamma  \rp^{\alpha - \ell} \gamma^{\ell} 
     \exp\lp\frac{-\ell \kappa_j^2}{2\sigma^2}\rp
     \exp\lp \lp\frac{\kappa_j\ell}{\sigma^2}\rp^2 \frac{1}{2}\sigma^2\rp\rp\\
     & = \frac{1}{\alpha -1} \log\lp 
     \sum_{\ell=0}^\alpha { \alpha \choose \ell}\lp 1-\gamma  \rp^{\alpha - \ell} \gamma^{\ell} 
     \exp\lp\frac{(\ell^2-\ell) \kappa_j^2}{2\sigma^2}\rp\rp
\end{align*}
where (a) is due to the generating function of normal distribution.

As a result, summing $j \in [d]$ yields
\begin{align*}
      \varepsilon^*\lp \alpha \rp 
     & \leq \max_{\kappa: \lV \kappa \rV_2 \leq \Delta_2, \lV \kappa \rV_\infty \leq \Delta_\infty } 
     \sum_j 
      \frac{1}{\alpha -1} \underbrace{\log\lp 
     \sum_{\ell=0}^\alpha { \alpha \choose \ell}\lp 1-\gamma  \rp^{\alpha - \ell} \gamma^{\ell} 
     \exp\lp\frac{(\ell^2-\ell) \kappa_j^2}{2\sigma^2}\rp\rp}_{\eqDef f(\kappa_j^2)}
\end{align*}
First, observe that (1) $\kappa^2 \mapsto f(\kappa^2)$ is increasing and convex (since it is log-sum-exp), and (2) $f(0) = 0$.
Next, define $\kappa_1^* \geq \kappa_2^* \geq \cdots \geq \kappa_d^*$ as the unique sequence such that
\begin{itemize}[leftmargin=1em,topsep=0pt]
    \item $\kappa_j^* = \Delta_\infty$ for any $j \leq \frac{\Delta^2_2}{\Delta^2_\infty}$;
    \item $\kappa_j^* = 0$ for any $j > \frac{\Delta^2_2}{\Delta^2_\infty} + 1$;
    \item $\sum_j (\kappa_j^*)^2 = \Delta_2^2$.
\end{itemize}
Then, it is obvious that $\lp\kappa_1^*\rp^2, \lp\kappa_2^*\rp^2, \cdots, \lp \kappa_d^*\rp^2$ is a majorization\footnote{See \url{https://en.wikipedia.org/wiki/Karamata\%27s_inequality} for a definition of ``majorization''.} of any $\kappa_1^2 \geq \kappa_2^2 \geq \cdots \geq \kappa_d^2$ such that $\sum_{j=1}^d \kappa_j^2 = \Delta_2$ and $\max_{j \in [d]} \kappa^2_j \leq \Delta_\infty^2$. Applying Karamata's inequality\footnote{\url{https://en.wikipedia.org/wiki/Karamata\%27s_inequality}} yields 
\begin{align*}
    & \max_{\kappa: \lV \kappa \rV_2 \leq \Delta_2, \lV \kappa \rV_\infty \leq \Delta_\infty } 
     \sum_j \frac{1}{\alpha -1} \log\lp 
     \sum_{\ell=0}^\alpha { \alpha \choose \ell}\lp 1-\gamma  \rp^{\alpha - \ell} \gamma^{\ell} 
     \exp\lp\frac{(\ell^2-\ell) \kappa_j^2}{2\sigma^2}\rp\rp\\
    & =  \sum_j \frac{1}{\alpha -1} \log\lp 
     \sum_{\ell=0}^\alpha { \alpha \choose \ell}\lp 1-\gamma  \rp^{\alpha - \ell} \gamma^{\ell} 
     \exp\lp\frac{(\ell^2-\ell) (\kappa_j^*)^2}{2\sigma^2}\rp\rp\\
    & = \frac{\lfloor \Delta_2^2/\Delta_\infty^2\rfloor}{\alpha -1} \log\lp 
     \sum_{\ell=0}^\alpha { \alpha \choose \ell}\lp 1-\gamma  \rp^{\alpha - \ell} \gamma^{\ell} 
     \exp\lp\frac{(\ell^2-\ell) \Delta_\infty^2}{2\sigma^2}\rp\rp\\
    & \quad+ \frac{1}{\alpha -1} \log\lp 
     \sum_{\ell=0}^\alpha { \alpha \choose \ell}\lp 1-\gamma  \rp^{\alpha - \ell} \gamma^{\ell} 
     \exp\lp\frac{(\ell^2-\ell) (\Delta_2^2 - \Delta_\infty^2\cdot \lfloor \Delta_2^2/\Delta_\infty^2\rfloor)}{2\sigma^2}\rp\rp\\
    & \leq \frac{ \Delta_2^2/\Delta_\infty^2}{\alpha -1} \log\lp 
     \sum_{\ell=0}^\alpha { \alpha \choose \ell}\lp 1-\gamma  \rp^{\alpha - \ell} \gamma^{\ell} 
     \exp\lp\frac{(\ell^2-\ell) \Delta_\infty^2}{2\sigma^2}\rp\rp,
\end{align*}
where the last inequality holds due to the convexity and the following Jensen's inequality:
 \begin{align*}
     & \frac{1}{\alpha -1} \log\lp 
      \sum_{\ell=0}^\alpha { \alpha \choose \ell}\lp 1-\gamma  \rp^{\alpha - \ell} \gamma^{\ell} 
      \exp\lp\frac{(\ell^2-\ell) (\Delta_2^2 - \Delta_\infty^2\cdot \lfloor \Delta_2^2/\Delta_\infty^2\rfloor)}{2\sigma^2}\rp\rp\\
      & = \frac{1}{\alpha -1} \log\lp 
      \sum_{\ell=0}^\alpha { \alpha \choose \ell}\lp 1-\gamma  \rp^{\alpha - \ell} \gamma^{\ell} 
      \exp\lp\frac{(\ell^2-\ell) \lp \Delta_\infty^2\lp  \Delta_2^2/\Delta_\infty^2- \lfloor \Delta_2^2/\Delta_\infty^2\rfloor\rp\rp}{2\sigma^2}\rp\rp\\
     & \leq \frac{1- \lp \Delta_2^2/\Delta_\infty^2 - \lfloor \Delta_2^2/\Delta_\infty^2\rfloor\rp}{\alpha -1} \log\lp 
      \sum_{\ell=0}^\alpha { \alpha \choose \ell}\lp 1-\gamma  \rp^{\alpha - \ell} \gamma^{\ell} 
      \exp\lp\frac{(\ell^2-\ell) \Delta_\infty^2 \cdot 0}{2\sigma^2}\rp\rp\\
      &\quad +\frac{\Delta_2^2/\Delta_\infty^2 - \lfloor \Delta_2^2/\Delta_\infty^2\rfloor}{\alpha -1} \log\lp 
      \sum_{\ell=0}^\alpha { \alpha \choose \ell}\lp 1-\gamma  \rp^{\alpha - \ell} \gamma^{\ell} 
      \exp\lp\frac{(\ell^2-\ell) \Delta_\infty^2\cdot 1}{2\sigma^2}\rp\rp \\
      & = \frac{\Delta_2^2/\Delta_\infty^2 - \lfloor \Delta_2^2/\Delta_\infty^2\rfloor}{\alpha -1} \log\lp 
      \sum_{\ell=0}^\alpha { \alpha \choose \ell}\lp 1-\gamma  \rp^{\alpha - \ell} \gamma^{\ell} 
      \exp\lp\frac{(\ell^2-\ell) \Delta_\infty^2}{2\sigma^2}\rp\rp.
 \end{align*}

This establishes the theorem.
\qedwhite

\subsection{Proof of Lemma~\ref{lemma:PG_PH_rdp_bound}}
To upper bound $D_{\alpha}\lp P_{\mb{P}|\mb{G}} \middle \Vert P_{\mb{P}|\mb{H}} \rp$, observe that for any coordinate $i \in [d]$, $\mb{P}_i \eqDef (\bm{p}^{(1)}_i, ..., \bm{p}^{(T)}_i)$ depends solely on $\bm{g}^{(1)}_i$, $\mb{s}^{(1)}_i$ and $\lp \mb{Z}^{(1)}_i,...,\mb{Z}^{(T)}_i \rp$. Therefore,
\begin{align*}
    D_{\alpha}\lp P_{\mb{P}|\mb{G}} \middle \Vert P_{\mb{P}|\mb{H}} \rp
    & = \sum_{i=1}^d D_{\alpha}\lp P_{\mb{P}_i|\mb{G}_i} \middle \Vert P_{\mb{P}_i|\mb{H}_i}\rp \\
    & = \sum_{i=1}^d D_{\alpha}\lp P_{\mb{p}^{(1)}_i, ..., \mb{p}^{(T)}_i|\mb{G}_i} \middle \Vert P_{\mb{p}^{(1)}_i, ..., \mb{p}^{(T)}_i|\mb{H}_i}\rp.
\end{align*}
Then, we claim that releasing $\lbp\bm{p}^{(t)}_i = \mb{M}_{t1}\lp \bm{g}^{(1)}_i \cdot \bm{S}^{(1)}_i \rp + \bm{Z}^{(t)}_i, t \in [T]\rbp$ is indeed an instance of (non-adaptive) subsampled Gaussian mechanism. By writing it in a vector form
\begin{align}
    \begin{bmatrix}
        \bm{p}^{(1)}_i\\
        \bm{p}^{(2)}_i\\
        \vdots\\
        \bm{p}^{(T)}_i
    \end{bmatrix}
    = \bm{g}^{(1)}_i\cdot\bm{S}^{(1)}_i\cdot
    \begin{bmatrix}
        \mb{M}_{11}\\
        \mb{M}_{21}\\
        \vdots\\
        \mb{M}_{T1}
    \end{bmatrix}
    + \begin{bmatrix}
        \mb{Z}^{(1)}_i\\
        \mb{Z}^{(2)}_i\\
        \vdots\\
        \mb{Z}^{(T)}_i
    \end{bmatrix},
\end{align}
it becomes clear as a subsampled Gaussian mechanism with sensitivity $\xi\lp \mb{M} \rp\cdot \lba \bm{g}^{(1)}_i \rba$. Since $\mb{M} = \mb{Q}\cdot\mb{C}$ and that $\mb{Q}$ is orthonormal, we have $\xi\lp \mb{M} \rp= \xi\lp \mb{C}\rp$. Also, by the geometrical assumption of data matrix $\mb{G}$, it holds that $\sum_{i=1}^d \lba \bm{g}^{(1)}_i \rba^2 \leq \Delta_2$ and $\lba \bm{g}^{(1)}_i \rba \leq \Delta_\infty$ for all $i$. Summing across $i \in [d]$ and applying Theorem~\ref{thm:mean_estimation_l2} yield
\begin{align}
    D_{\alpha}\lp P_{\mb{P}|\mb{G}} \middle \Vert P_{\mb{P}|\mb{H}} \rp \leq \frac{ \kappa_2^2/\kappa_\infty^2}{\alpha-1} \log\lp (1-\gamma)^{\alpha-1}\lp \gamma(\alpha-1) + 1 \rp + \sum_{\ell=1}^\alpha { \alpha \choose \ell}\lp 1-\gamma  \rp^{\alpha - \ell} \gamma^{\ell} e^{(\ell-1) \alpha\frac{\kappa_\infty^2}{2\sigma^2}} \rp,
\end{align}
establishing the desired result.